# Deformation of the three-term recursion relation and the generation of new orthogonal polynomials

A. D. Alhaidari

*Physics Department, King Fahd University of Petroleum & Minerals, Box 5047,*

*Dhahran 31261, Saudi Arabia*

E-mail: haidari@mailaps.org

We find solutions for a linear deformation of the symmetric three-term recursion relation. The orthogonal polynomials of the first and second kind associated with the deformed relation are obtained. The new density (weight) function is written in terms of the original one and the deformation parameters.

**Introduction**: The use of orthogonal polynomials in the analytic solution of various problems in physics is overwhelming. The reason behind the remarkable presence of these objects may lie in one of the fundamental problems of theoretical physics – the solution of the eigenvalue problem $(H-x)|\chi(x)\rangle = 0$, where $H$ is a hermitian operator and $x$ is real. The solution of this equation for a general observable $H$ of a given physical system is often very difficult to obtain. However, for simple systems or for those with high degree of symmetry, an analytic solution is feasible. On the other hand, for a large class of problems that model realistic physical systems, the operator $H$ could be written as the sum of two components: $H = H_0 + \tilde{V}$. The "reference" operator $H_0$ is often simpler and carries a high degree of symmetry while $\tilde{V}$ is not, but it is usually endowed with either one of two properties. Its contribution is either very small compared to $H_0$ or is limited to a finite region in configuration or function space. Perturbation techniques are used to give a numerical evaluation of its contribution in the former case, while algebraic methods are used in the latter [1]. Thus, the analytic problem is confined to finding the solution of the reference $H_0$–problem

$$(H_0 - x)|\psi(x)\rangle = 0 \qquad (1)$$

Due to the higher degree of symmetry of this problem, it is frequently possible to find a special basis for the solution space of this equation that supports a tridiagonal matrix representation for $H_0$. Let $\{\phi_n\}_{n=0}^{\infty}$ be such a basis, which is complete, orthogonal and belongs to the space of square integrable functions. Therefore, we can write the matrix representation of $H_0$ in this basis as

$$H_0 = \begin{pmatrix} a_0 & b_0 & & & & & \\ b_0 & a_1 & b_1 & & & 0 & \\ & b_1 & a_2 & b_2 & & & \\ & & b_2 & \times & \times & & \\ & & & \times & \times & \times & \\ & 0 & & & \times & \times & \times \\ & & & & & \times & \times \end{pmatrix} \quad (2)$$

If we expanded the reference wave function as $|\psi(x)\rangle = \sum_n d_n(x)|\phi_n\rangle$ then the equivalent matrix representation of the eigenvalue equation (1) gives the following symmetric three-term recursion relation

$$xd_n(x) = a_n d_n(x) + b_{n-1} d_{n-1}(x) + b_n d_{n+1}(x); \qquad n \geq 1 \qquad (3)$$

This is one way of illustrating the intimate relation between the symmetry of a physical system, whose dynamics is described by $H_0$ in (2), and the theory of orthogonal polynomials associated with this recursion relation [2]. Another approach is found in the studies that investigate the properties of tridiagonal matrices and their relations to quadrature approximation, continued fractions, and the theory of orthogonal polynomials [3,4]. In this article we want to explore the extent of this relation in the case where the physical system is being subjected to alterations (deformations) of its dynamics. We will show, specifically, that this relation will persist despite a particular kind of one- and three-parameter linear deformations.

Typically, there are two solutions to the recursion relation (3). One results in a regular wave function and the other doesn't. The coefficients of the regular one, $d_n(x) = p_n(x)$, satisfy a homogeneous initial relation. However, the coefficients of the irregular solution, $d_n(x) = q_n(x)$, satisfy an inhomogeneous one. These initial relations that complement the above recursion, for $n = 0$, are

$$xp_0 = a_0 p_0 + b_0 p_1$$
$$xq_0 = a_0 q_0 + b_0 q_1 - 1$$

with $p_0(x) = 1$ and $q_0(x) = 0$. $p_n(x)$ is a polynomial of degree $n$ and named "polynomial of the first kind", while $q_n(x)$ is a polynomial of degree $n-1$ and named "polynomial of the second kind"[4]. They also satisfy a Wronskian-like relation (Liouville-Ostrogradskii formula) that reads

$$b_{n-1}\left[p_{n-1}(x)q_n(x) - p_n(x)q_{n-1}(x)\right] = 1 ; \quad n \geq 1$$

The resolvent operator (Green's function) for this system is formally defined for any complex number $z$ by $G(z) \equiv (H_0 - z)^{-1}$. It has point singularities (respectively, branch cut) on the real line corresponding to the discrete (continuous) spectrum of $H_0$. Its (0,0) component can simply be written as the limit of the polynomial ratio

$$G_{00}(z) = -\lim_{n\to\infty}\left[q_n(z)/p_n(z)\right] \tag{4}$$

and has the following continued fraction representation[3-5]

$$G_{00}(z) = -\left\{z - a_0 - b_0^2\left[z - a_1 - b_1^2\left(z - a_2 - b_2^2(...)^{-1}\right)^{-1}\right]^{-1}\right\}^{-1} \tag{5}$$

The density (weight) function $\rho(x)$ associated with these polynomials which appears in the orthogonality relation $\int \rho(x) p_n(x) p_m(x) dx = \delta_{nm}$ can be written as

$$\rho(x) = \frac{1}{\pi}\text{Im}\left[G_{00}(x + i0)\right] \tag{6}$$

**The Deformation**: We choose to carry out the development using an alternative notation based on the following set of two-component vector polynomials

$$P_n \equiv \begin{pmatrix} p_n \\ p_{n+1} \end{pmatrix}, \quad Q_n \equiv \begin{pmatrix} q_n \\ q_{n+1} \end{pmatrix}$$

In this notation, the recursion relation (3) could be rewritten as

$$D_n = \begin{pmatrix} 0 & 1 \\ -\dfrac{b_{n-1}}{b_n} & \dfrac{x-a_n}{b_n} \end{pmatrix} D_{n-1} \equiv \Im_n D_{n-1}; \qquad n \geq 1 \qquad (3')$$

where $D_n$ stands for either $P_n$ or $Q_n$. The initial vectors are

$$P_0 \equiv \begin{pmatrix} 1 \\ (x-a_0)/b_0 \end{pmatrix} \text{ and } Q_0 \equiv \begin{pmatrix} 0 \\ 1/b_0 \end{pmatrix}$$

Now, we introduce a one-term linear deformation of the recursion (3) or, equivalently (3'), as the mapping

$$a_0 \to \hat{a}_0 = a_0 + \mu \qquad (7)$$

where $\mu$ is a real constant parameter. This is equivalent to the transformation

$$\left(\hat{H}_0\right)_{nm} = \left(H_0\right)_{nm} + \mu \delta_{n0} \delta_{m0} \qquad (8)$$

A physical interpretation of this deformation could be given if $H_0$ were to represent the dynamics of the given system. For such system, this deformation could be considered as a model for one-term separable potential coupling [6]. Anyhow, this deformation induces the following change in the orthogonal polynomials so that they form a new complete set of solutions for the deformed recursion relation

$$\hat{P}_n = P_n - \mu Q_n$$
$$\hat{Q}_n = Q_n$$

Moreover, using the continued fraction representation of $G_{00}(z)$ in (5) we get the following expression for the deformed resolvent $\hat{G}_{00}(z) = G_{00}(z)/[1+\mu G_{00}(z)]$ giving the deformed density function as

$$\hat{\rho}(x) = \rho(x)/|1 + \mu G_{00}(x+i0)|^2 \qquad (9)$$

The three-parameter deformation, on the other hand, is defined by

$$\hat{H}_0 = H_0 + \begin{pmatrix} \mu_+ & \mu_0 \\ \mu_0 & \mu_- \end{pmatrix} \qquad (10)$$

where $\mu_\pm, \mu_0$ are real. It is equivalent to the map

$$a_0 \to \hat{a}_0 = a_0 + \mu_+,\ a_1 \to \hat{a}_1 = a_1 + \mu_-,\ \text{and } b_0 \to \hat{b}_0 = b_0 + \mu_0 \qquad (11)$$

which generates the following polynomial transformations

$$\hat{P}_n = P_n - \frac{b_0}{b_0 + \mu_0}\left[\mu_+ + \frac{\mu_0}{b_0}(x - a_0)\right]Q_n + \frac{\mu_-/b_1}{b_0 + \mu_0}\left[\mu_+ - \frac{\mu_0}{\mu_-}(b_0 + \mu_0) + a_0 - x\right]\tilde{P}_{n-2}$$

$$\hat{Q}_n = \frac{b_0}{b_0 + \mu_0} Q_n - \frac{\mu_-/b_1}{b_0 + \mu_0}\tilde{P}_{n-2}$$

where $\tilde{P}_n = P_n(a_n \to a_{n+2}, b_n \to b_{n+2})$ and $\tilde{P}_{-1} \equiv \binom{0}{1}$, $\tilde{P}_{-2} \equiv \binom{0}{0}$. These polynomials, which are called the "abbreviated polynomials,"[4] satisfy the recursion relation $\tilde{P}_n = \mathfrak{I}_{n+2}\tilde{P}_{n-1}$, where the 2×2 matrix $\mathfrak{I}_n$ is defined in the recursion relation (3') above. They are associated with a tridiagonal matrix obtained from $H_0$ in (2) by deleting the first two rows and first two columns. They are written in terms of the original polynomials as follows[4]

$$\tilde{P}_n = -\frac{b_1}{b_0}\left[P_{n+2} + (a_0 - x)Q_{n+2}\right]; \qquad n \geq 0$$

The deformed resolvent operator can now be written in terms of the original one as

$$\hat{G}_{00}(z) = -\left\{z - a_0 - \mu_+ + (b_0 + \mu_0)^2\left[\mu_- - b_0^2\left(z - a_0 + G_{00}^{-1}(z)\right)^{-1}\right]^{-1}\right\}^{-1}$$

resulting in the deformed density function

$$\hat{\rho}(x) = \left\|\left[\frac{1 + (x - a_0)G_{00}(x)}{b_0(b_0 + \mu_0)}\right]\left\{(b_0 + \mu_0)^2 + (x - a_0 - \mu_+)\left[\mu_- - \frac{b_0^2 G_{00}(x)}{1 + (x - a_0)G_{00}(x)}\right]\right\}\right\|^{-2} \rho(x) \quad (12)$$

where $G_{00}(x) \equiv G_{00}(x + i0)$.

Higher order linear deformation could be pursued following the same formalism presented above for the one- and three-parameter deformations. In each order $N$ the number of deformation parameters is $N(N+1)/2$.

**Example**: As an example we consider the case where the recursion coefficients are $a_n = 0, b_n = 1/2$. These will result in a one-band density whose non-vanishing support is the real interval with end points $a_\infty \pm 2b_\infty = \pm 1$ [7]. The three-term symmetric recursion relation associated with these coefficients reads $2x d_n(x) = d_{n-1}(x) + d_{n+1}(x)$. The regular solution of this recursion is the well-known Chebyshev polynomials [8], which could be written in any one of several alternative forms of which we choose the following

$$p_n(x) = (n+1)\,_2F_1(-n, n+2; 3/2; \tfrac{1-x}{2})$$

where $_2F_1(a,b;c;z)$ is the Gauss hypergeometric function. The irregular polynomials in this case are simple and could be written as $q_n(x) = 2 p_{n-1}(x)$ with $p_{-1}(x) \equiv 0$. The (0,0) component of the resolvent operator can simply be obtained using the continued fraction representation (5) giving $G_{00}(z) = -2z + 2\sqrt{z^2 - 1}$. Thus, the density function, which is obtained from this $G_{00}(z)$ using relation (6), is

$$\rho(x) = \frac{2}{\pi}\sqrt{1 - x^2}\,; \qquad x \in [-1, +1]$$

The one-term deformation defined in (7) or, equivalently, (8) produces the following deformed orthogonal polynomials

$$\hat{p}_n(x) = p_n(x) - 2\mu p_{n-1}(x)$$
$$\hat{q}_n(x) = 2 p_{n-1}(x)$$

Using (9) we obtain the deformed density function

$$\hat{\rho}(x) = \frac{\rho(x)}{1+4\mu(\mu-x)}$$

On the other hand, the three-term deformation defined by (10) or (11) gives the following new orthogonal polynomials:

$$\hat{p}_0 = 1, \quad \hat{p}_1 = 2\frac{x-\mu_+}{1+2\mu_0}$$

$$\hat{q}_0 = 0, \quad \hat{q}_1 = \frac{2}{1+2\mu_0}$$

$$\hat{p}_n = p_n - 2\frac{\mu_+ + 2\mu_0 x}{1+2\mu_0} p_{n-1} + \frac{4\mu_-}{1+4\mu_0}\left[-\mu_+ + \frac{\mu_0}{\mu_-}(\mu_0 + \tfrac{1}{2}) + x\right](p_n - 2xp_{n-1})$$

$$\hat{q}_n = \frac{2}{1+2\mu_0} p_{n-1} + \frac{4\mu_-}{1+4\mu_0}(p_n - 2xp_{n-1})$$
; $n \geq 2$

Note that the combination $p_n - 2xp_{n-1}$ is a polynomial of degree $n$–2. The new density function is obtained in terms of the original one using (12) which gives

$$\hat{\rho}(x) = \left\{(1+2\mu_0)^2 + 4(\mu_+ - x)\left[x - 2\mu_- + (\mu_+ - x)/(1+2\mu_0)^2\right]\right\}^{-1} \rho(x) \quad (13)$$

It is instructive to compare the analytic form of this density function with those that could be obtained independently by any number of possible numerical schemes. In reference [9] three numerical methods were presented to extract approximate, yet highly accurate, density-of-state information over a continuous range of energies from a finite symmetric Hamiltonian matrix. Starting with the (finite) tridiagonal deformed matrix $\hat{H}_0$ in (10) these methods will be used to evaluate the density function, which will be compared with values obtained from the analytic form (13). The first ("Analytic Continuation") method relies on the analytic continuation of the finite polynomial ratio approximation (4) of the resolvent operator. The other two ("Dispersion Correction" and "Stieltjes Imaging") methods are based on the fact that the density function is related to the distribution of the eigenvalues of the finite matrix and one of its submatrices. Figure 1 shows the results of using these three methods with the parameters given in the caption. The agreement with the analytic result is excellent despite the relatively low matrix dimension.

**FIGURE CAPTION:**

FIG. 1: Graphs of the density function (solid curve) as given by the analytic expression in equation (13) for the three-parameter deformation of the Chebyshev polynomials superimposed by the numerical evaluation ("+" points). The "Analytic Continuation", "Dispersion Correction", and "Stieltjes Imaging" methods of reference [9] were used for the numerical evaluation of the density in figures (a), (b), and (c), respectively. The dimension of the deformed matrix $\hat{H}_0$ in the numerical calculations was set to 10. The deformation parameters were assigned the following values $\mu_+ = 0.2, \mu_- = 0.0, \mu_0 = -0.1$

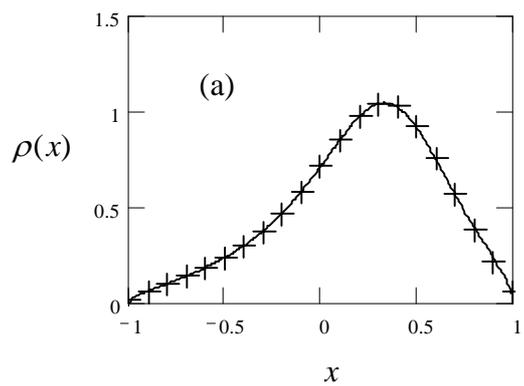

Fig. 1(a)

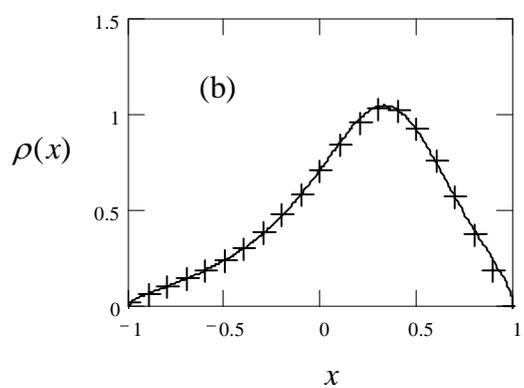

Fig. 1(b)

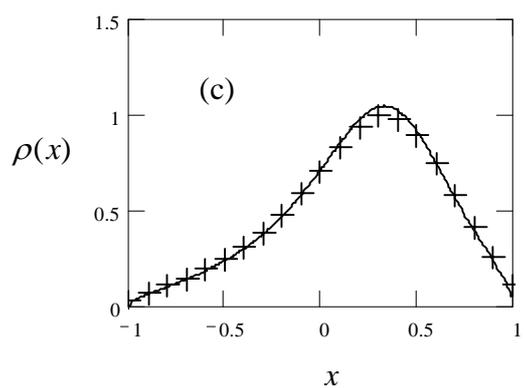

Fig. 1(c)

A. D. Alhaidari
Journal of Mathematical Physics

12